\documentclass[conference]{IEEEtran}
\usepackage{cite}
\usepackage{amsmath,amssymb,amsfonts}
\usepackage{algorithmic}
\usepackage{graphicx}
\usepackage{textcomp}
\usepackage{xcolor}
\usepackage{stfloats} 
\usepackage{amsmath}
\usepackage{afterpage}
\def\BibTeX{{\rm B\kern-.05em{\sc i\kern-.025em b}\kern-.08em
    T\kern-.1667em\lower.7ex\hbox{E}\kern-.125emX}}

\begin{document}

\title{Lesion Segmentation in FDG-PET/CT Using Swin
Transformer U-Net 3D: A Robust Deep Learning
Framework\\
}

\author{\IEEEauthorblockN{1\textsuperscript{st} Shovini Guha}
\IEEEauthorblockA{\textit{Computer Science and Engineering (Artificial Intelligence and Machine Learning (of Aff.)} \\
\textit{Institute of Engineering and Management (of Aff.)}\\
Kolkata, India \\
shoviniguha@gmail.com}
\and
\IEEEauthorblockN{2\textsuperscript{nd} Dwaipayan Nandi}
\IEEEauthorblockA{\textit{Computer Science and Engineering (Artificial Intelligence and Machine Learning (of Aff.)} \\
\textit{Institute of Engineering and Management (of Aff.)}\\
Kolkata, India \\
nandidwaipayan@gmail.com}
}

\maketitle

\begin{abstract}
Accurate and automated lesion segmentation in Positron Emission Tomography / Computed Tomography (PET/CT) imaging is essential for cancer diagnosis and therapy planning. This paper presents a Swin Transformer UNet-3D (SwinUNet3D) framework for lesion segmentation in Fluorodeoxyglucose Positron Emission Tomography / Computed Tomography (FDG-PET/CT) scans. By combining shifted window self-attention with U-Net style skip connections, the model captures both global context and fine anatomical detail.

We evaluate SwinUNet3D on the AutoPET III FDG dataset and compare it against a baseline 3D U-Net. Results show that SwinUNet3D achieves a Dice score of 0.88 and IoU of 0.78, surpassing 3D U-Net (Dice 0.48, IoU 0.32) while also delivering faster inference times. Qualitative analysis demonstrates
improved detection of small and irregular lesions, reduced false
positives, and more accurate PET/CT fusion.

While the framework is currently limited to FDG scans and trained under modest GPU resources, it establishes a strong foundation for future multi-tracer, multi-center evaluations and benchmarking against other transformer-based architectures. Overall, SwinUNet3D represents an efficient and robust approach to PET/CT lesion segmentation, advancing the integration of transformer-based models into oncology imaging workflows.

\end{abstract}


\section{Introduction}
Medical imaging plays a central role in modern oncology, supporting early diagnosis, staging, and therapy monitoring. Among available modalities, positron emission tomography combined with computed tomography (PET/CT)\cite{b17} has become a clinical standard, providing both functional and anatomical information within a single scan. Fluorodeoxyglucose (FDG)-PET\cite{b15} highlights regions of elevated glucose metabolism—often indicative of malignancy—while CT provides high-resolution anatomical context. Together, FDG PET/CT enables clinicians to detect, localize, and monitor tumors with high sensitivity and specificity.

A critical step in PET/CT analysis is lesion segmentation\cite{b10}, i.e., delineating tumor regions from surrounding tissues. Traditionally, segmentation has been performed manually by radiologists and nuclear medicine experts.While considered the gold standard, manual delineation is labor-intensive, subjective, and prone to inter- and intra- observer variability. These challenges are particularly limiting in large-scale studies and high throughput clinical workflows, where consistency and speed are essential. Automated segmentation methods\cite{b15} are therefore needed to support reproducible and efficient decision-making in oncology.

Over the past decade, deep learning has transformed medical image analysis. Convolutional Neural Networks (CNNs)\cite{b1}, particularly U-Net\cite{b15} and its 3D variants\cite{b7}, have demonstrated strong performance in segmentation tasks by combining hierarchical feature extraction with encoder–decoder structures and skip connections. In PET/CT, 3D U-Nets have shown promise for capturing volumetric tumor information. However, CNN-based models remain limited in their ability to model long-range spatial dependencies, which are essential for detecting small, dispersed, or irregular lesions.Prior studies\cite{b28} have shown that anatomy-aware and tracer-specific CNN models can improve robustness in multi-center PET/CT segmentation.

Recently, transformer-based architectures\cite{b16} have emerged as powerful alternatives to CNNs in vision tasks. The Swin Transformer\cite{b13}, in particular, introduces shifted window-based self-attention\cite{b19}, enabling efficient modeling of both local and global dependencies with reduced computational overhead. Its hierarchical design makes it naturally compatible with UNet-like structures for segmentation. In medical imaging,
Swin-based models have begun to outperform conventional CNNs, particularly in scenarios where contextual reasoning is crucial.

Building on this progress, we propose a Swin Transformer U-Net 3D (SwinUNet3D) framework for automated lesion segmentation in FDG-PET/CT imaging. The model leverages dual-modality fusion (PET for functional activity, CT for anatomical context) and incorporates 3D Swin Transformer blocks within a U-Net architecture. This design allows the network to capture both local detail and long-range dependencies in volumetric data while maintaining computational efficiency.

The contributions of this paper are threefold:
\begin{enumerate} 
    \item We adapt the Swin Transformer U-Net 3D for dual-channel FDG-PET/CT segmentation.
    \item We propose a preprocessing pipeline to standardize volumetric PET/CT data.
    \item We evaluate the framework on the autoPET III\cite{b3} dataset, achieving competitive results and demonstrating potential for clinical translation.
\end{enumerate}

This work represents a step toward clinically viable transformer-based segmentation frameworks, addressing the need for reproducible, automated lesion analysis in oncology imaging.

\section{Methodology}
This section outlines the experimental framework used to implement and evaluate the Swin Transformer U-Net 3D (SwinUNet3D) model for lesion segmentation in FDG-PET/CT images. We begin by describing the acquisition and preprocessing of the dataset, including normalization and dual-channel
construction. Next, we present the design and architecture of the SwinUNet3D model, highlighting the integration of transformer-based self-attention\cite{b19} within a U-Net\cite{b15} structure. We then detail the training configuration, including loss functions, optimization strategy, and training parameters.
Finally, we describe the evaluation metrics and validation procedures used to assess segmentation performance.
\subsection{Data Acquisition and Preprocessing}
\subsubsection{Dataset Description}
We used the FDG-PET/CT dataset from University Hospital T¨ubingen, released as part of the AutoPET III challenge\cite{b3}. Each case consists of co-registered
PET and CT volumes with binary expert lesion masks. The cohort includes 501 patients with confirmed malignancies(melanoma, lymphoma, or lung cancer) and 513 negative controls, reflecting substantial clinical heterogeneity.
\subsubsection{Preprocessing Steps}
The original FDG-PET/CT image volumes used in this study were acquired from the autoPET III challenge dataset, where each case includes co-registered
PET and CT volumes with a spatial resolution of 400 × 400 pixels in the axial plane and a variable number of slices(depth) across patients. In order to make the dataset compatible with the SwinUNet3D training pipeline and to improve model performance and convergence, the following preprocessing steps were applied as illustrated in Fig. 1 :
\paragraph{Input Initialization}
PET and CT volumes for each patient are read as co-registered 3D arrays. These volumes are structured as a 2-channel input tensor with fixed height and width of 400 × 400 and a variable number of slices (depth) depending on the scan.
\paragraph{Normalization}
Each PET and CT image volume is normalized such that pixel intensities are scaled between 0 and 1, with the maximum intensity value set to 1. PET images, in particular, contain wide dynamic ranges in SUV values. Normalizing ensures numerical stability during training and prevents any one modality or region from dominating the gradients. It also improves convergence speed and helps the model generalize better by removing intensity scale variance between scans.
\paragraph{Padding}
If the number of slices (depth) in a volume is not divisible by 16, blank slices (zeros) are added at the end of the volume until the depth becomes a multiple of 16.

SwinUNet3D uses a hierarchical encoder-decoder structure with multiple downsampling and upsampling layers. To maintain symmetry and dimensional consistency during these operations, the depth must be divisible by the overall scaling factor (which is 16 in this case). Padding avoids dimension mismatches and ensures smooth propagation through all transformer blocks and pooling layers.
\paragraph{Stacking}
The normalized PET and CT volumes from a patient are stacked along the channel axis, producing a dual-channel input of shape: [Height x Width x Depthx Channels] = 400 × 400 × D × 2 PET provides metabolic activity information, while CT captures anatomical structure. Stacking both channels
allows the model to jointly learn complementary features from both modalities. This multi-modal fusion enables better boundary delineation and lesion localization compared to single-modality inputs.
\paragraph{Patching}
The full volume is divided into overlapping or non-overlapping 3D patches of size 400 × 400 × 16 × 2 (H × W × D × C). This reduces GPU memory consumption, allowing for mini-batch training. and improves the model’s ability to learn local 3D spatial patterns, especially in smaller lesion regions. It also serves as a form of data augmentation by exposing the model to multiple spatial contexts during training.

This preprocessing pipeline ensures architectural compatibility,
balanced multimodal fusion, and improved convergence.

\subsection{Model Design and Architecture}
The core of this study is the Swin Transformer U-Net 3D (SwinUNet3D) model — a hybrid architecture that effectively combines the strengths of Vision Transformers and U-Net\cite{b15} to perform accurate and robust 3D medical image segmentation. Specifically designed for volumetric data, SwinUNet3D processes dual-channel PET/CT inputs and is optimized for segmenting complex lesion structures from FDG-PET/CT scans.
\subsubsection{Input Format}
Each input volume is a 3D tensor with shape (2 × 16 × 400 × 400), representing:
\begin{itemize}
    \item 2 channels: PET and CT modalities
    \item 16 slices (depth)
    \item 400 × 400 pixels per slice (height and width)
\end{itemize}

This dual-modality input allows the model to learn from both
anatomical and functional imaging data simultaneously.

\begin{figure}
    \centering
    \includegraphics[width=1\linewidth]{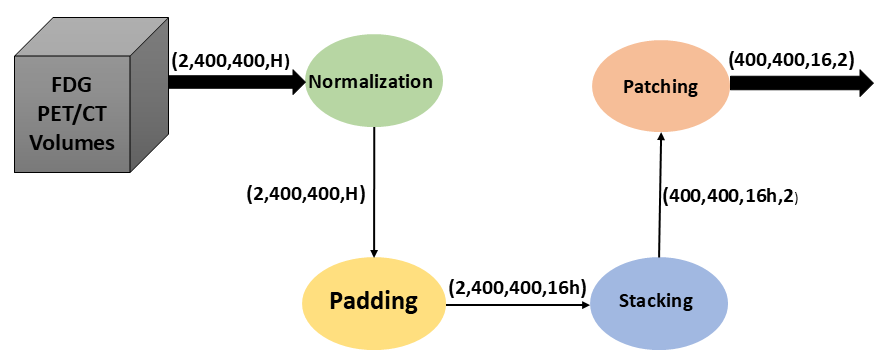}
    \caption{Preprocessing workflow applied to PET/CT data. Raw inputs undergo intensity normalization, zero-padding, and patching before being fed into the SwinUNet3D network. This ensures consistent input dimensions and efficient utilization of 3D context.}
    \label{fig:preproc}
\end{figure}

\subsubsection{Architecture Overview}
The SwinUNet3D architecture follows a symmetric encoder-decoder structure, similar to traditional U-Net, but replaces standard convolutions with Swin
Transformer blocks, which apply shifted window-based self-attention to capture both local and contextual dependencies in 3D volumes. Overall, the model contains 2 encoder blocks, 1 bottleneck block, and 2 decoder blocks, enabling it to balance local detail preservation with global context modeling
while maintaining computational efficiency through hierarchical window-based attention.

The model consists of the following key components:
\paragraph{Patch Embedding}
The Patch Embedding layer converts the 3D PET/CT input volume into a set of lower-dimensional feature tokens suitable for transformer processing. Implemented via a 3D convolution with kernel and stride equal to the patch size, it splits the volume into non-overlapping 3D patches and projects each patch into a feature space of specified dimension. This step reduces spatial resolution, preserves local context, and produces a tokenized representation compatible with Swin Transformer blocks. In SwinUNet3D, patch embedding is essential for efficiently handling large volumetric inputs while enabling the model to capture both anatomical and functional details from the dual-channel PET/CT data.
\paragraph{Encoder Block}
Each encoder block in SwinUNet3D consists of two sequential Swin Transformer 3D blocks(SwinBlock3D). A SwinBlock3D first normalizes its input, partitions it into local 3D windows, and applies multi-head self-attention to model relationships within each window. In alternating blocks, a shifted window strategy is used to enable cross-window information exchange, allowing the network to capture broader spatial dependencies without global attention. After attention, a multi-layer perceptron (MLP) refines the features before residual connections merge them back into the main stream. By stacking two such blocks, each encoder stage progressively enriches feature representations, balancing local detail capture with contextual awareness before downsampling to the next stage.
\paragraph{Downsampling Block}
The downsampling block in SwinUNet3D reduces the spatial dimensions (depth, height, and width) of the feature maps by a factor of 2 while increasing the number of channels. It is implemented as a 3D convolution with a kernel size of 2 and stride of 2. This operation serves two purposes: by reducing spatial resolution, subsequent layers can focus on more abstract, high-level features and lower resolution feature maps reduce memory usage and computation
cost in deeper layers. In the encoder path, downsampling is applied after each encoder block to progressively build a multi-scale representation of the input PET/CT volumes.
\paragraph{Bottleneck Block} 
The bottleneck block in Swin-UNet3D sits at the deepest point of the network, between the encoder and decoder paths. At this stage, the feature maps have the smallest spatial dimensions but the highest channel depth, allowing the network to capture global context and high-level semantic features. The bottleneck consists of two consecutive Swin Transformer 3D blocks, which use shifted window-based self-attention to model long-range dependencies across the entire feature map. This enables the model to integrate information from distant regions in the volume—critical for identifying lesions that may be small, irregular, or dispersed—before passing the enriched representation to the decoder for reconstruction.
\paragraph{Decoder Block} 
The decoder block in SwinUNet3D is responsible for reconstructing the segmentation map from the compressed feature representation learned by the encoder. It begins with upsampling to double the spatial dimensions of the input features. These upsampled features are then concatenated with the corresponding encoder features via skip connections, which reintroduce fine-grained spatial details lost during downsampling. The combined features pass through two Swin Transformer 3D blocks—the same type used in the encoder—to refine and integrate local detail with contextual information. This process is repeated at each decoder stage, gradually restoring the full resolution of the input while maintaining the rich semantic context learned at deeper levels.
\paragraph{Upsampling Block} 
The upsampling block in Swin-UNet3D restores the spatial resolution of feature maps by a factor of 2 while reducing the number of channels, enabling reconstruction of high-resolution segmentation outputs. This is typically achieved using trilinear interpolation or transposed 3D convolution, followed by feature fusion with the corresponding encoder output through skip connections. The restored feature maps allow the decoder to recover fine anatomical details lost during downsampling, ensuring precise
lesion boundary delineation in the final segmentation.
\paragraph{Skip Connections}
Classic U-Net-style skip connections are used between encoder and decoder blocks to preserve spatial detail lost during downsampling. These connections help in localizing lesion boundaries accurately.
\paragraph{Segmentation Head}
The decoder output is passed through a 1 × 1 × 1 convolutional layer to generate the final segmentation logits. The output tensor represents the voxel-wise lesion probability map, which is later thresholded to obtain the binary segmentation mask.

 \begin{figure*}[t] % you can use t (top), b (bottom), or h (here)
    \centering
    \includegraphics[width=\textwidth]{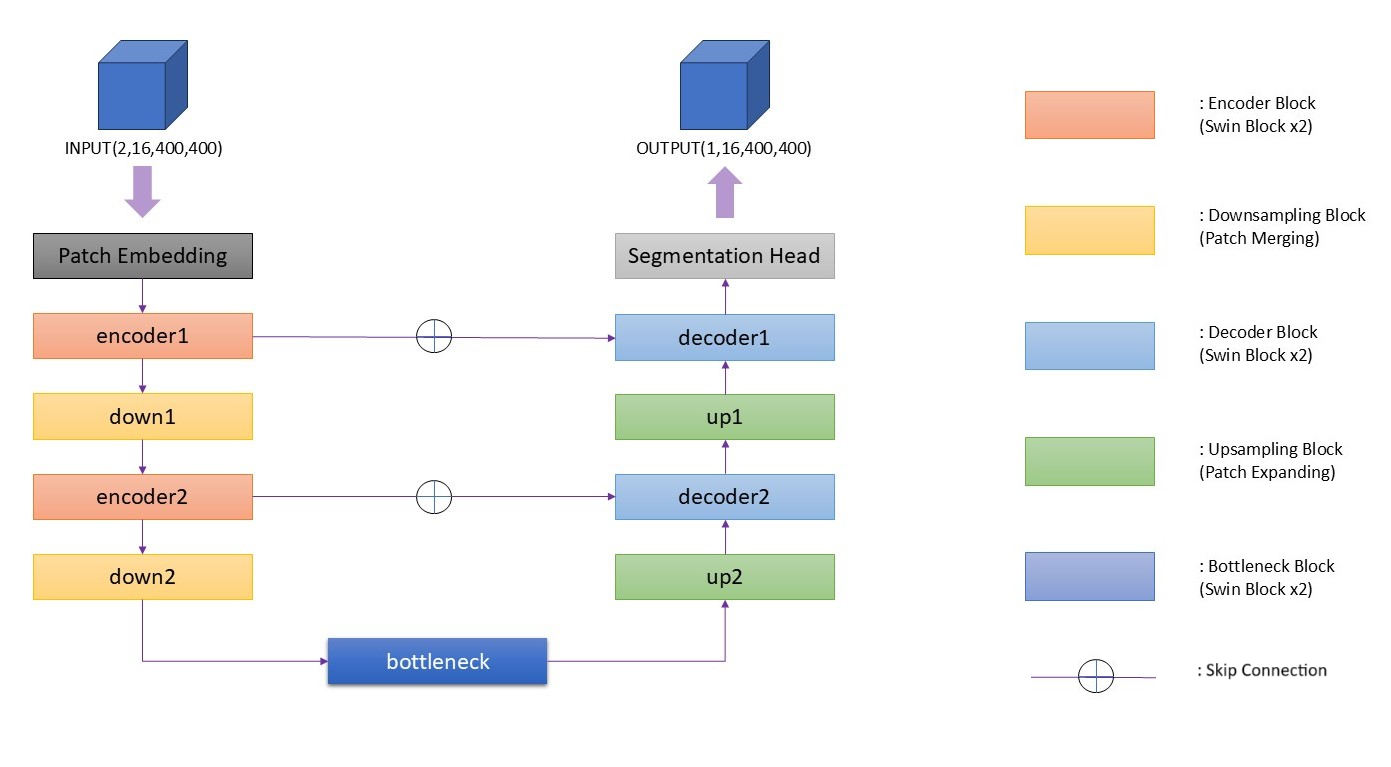}
    \caption{Overview of the proposed SwinUNet3D architecture. The model adopts a U-Net-like encoder–decoder design\cite{b15} with hierarchical Swin Transformer blocks\cite{b13}, patch embedding, bottleneck, and skip connections (see labels in the figure). This structure enables both local feature extraction and global context modeling for accurate 3D lesion segmentation.}
    \label{fig:architecture}
\end{figure*}

\paragraph*{Advantages of SwinUNet3D}

\begin{itemize}
    \item Efficient Attention Mechanism: Swin Transformers operate on local windows, avoiding the computational overhead of full self-attention in large 3D volumes. The use of shifted windows maintains global information flow while being memory-efficient.
    \item Contextual and Spatial Precision: By combining the global reasoning of transformers with spatial localization of U-Net-style skip connections, the architecture can effectively segment fine and complex structures in medical images.
    \item Generalization and Robustness: The hierarchical structure allows for multi-scale representation learning, which improves generalization across patient anatomy and tumor morphology.
\end{itemize}

SwinUNet3D serves as a powerful and scalable model for 3D medical image segmentation. Its attention-based design, paired with efficient patch embedding and skip-connected decoding, enables high segmentation accuracy even on challenging datasets like FDG-PET/CT, where lesion boundaries may be irregular or sparsely defined. This model represents a step forward in bridging transformer architectures with clinical imaging tasks.

\subsection{Training Configuration}
The SwinUNet3D model was implemented in PyTorch and
trained using supervised learning on paired PET/CT volumes
with binary segmentation masks. All experiments were conducted
on a system equipped with an NVIDIA RTX 3050
GPU.
\paragraph*{Training Setup}
\begin{itemize}
    \item Optimizer: Adam \cite{b26} with initial learning rate 1e-4.
    \item Batch Size: 2, constrained by GPU memory.
    \item Epochs: 100 with early stopping based on validation Dice score.
    \item Loss Function: Binary Focal Loss\cite{b11} ($\alpha = 0.25, \gamma = 2$) to mitigate class imbalance between lesion and background voxels.
    \item Regularization: Data normalization and residual connections\cite{b27} stabilized training under memory-limited
conditions.
\end{itemize}

\subsection{Evaluation and Validation}
To assess the segmentation performance of the proposed
SwinUNet3D model, we employ three complementary evaluation
metrics: Dice Similarity Coefficient (DSC)\cite{b20}\cite{b21},
Intersection over Union (IoU)\cite{b22}, and Binary Focal Loss.
These metrics jointly capture overlap quality, boundary accuracy,
and robustness to class imbalance, which are all critical
in lesion segmentation.

\begin{table}[h]
\centering
\caption{Model and Training Hyperparameters for SwinUNet3D}
\label{tab:hyperparameters}
\begin{tabular}{p{0.45\linewidth} p{0.45\linewidth}}
\hline
\textbf{Parameter} & \textbf{Value} \\
\hline
\multicolumn{2}{l}{\textit{Model Hyperparameters}} \\
Input channels & 2 (PET + CT) \\
Number of output classes & 1 \\
Base embedding dimension ($base\_dim$) & 32 \\
Window size & 2 \\
Patch size (H $\times$ W $\times$ D) & 16 $\times$ 400 $\times$ 400 \\
\hline
\multicolumn{2}{l}{\textit{Training Hyperparameters}} \\
Optimizer & Adam\cite{b26} \\
Learning rate & $1 \times 10^{-4}$ \\
Batch size & 2 \\
Epochs & 100 (early stopping on validation Dice) \\
Loss function & Binary Focal Loss\cite{b11} ($\alpha=0.25$, $\gamma=2$) \\
\hline
\textit{Total trainable parameters} & 810,721 \\
\hline
\end{tabular}
\end{table}

\paragraph*{Dice Score Coefficient}
The Dice Similarity Coefficient
(DSC), also known as the Sørensen–Dice index, quantifies the
overlap between predicted segmentation (P) and ground truth
(G) as:
\begin{equation}
\text{DSC} = \frac{2|P \cap G|}{|P| + |G|}
\end{equation}

DSC ranges from 0 (no overlap) to 1 (perfect overlap). It
is widely used in medical image segmentation because it
directly measures region overlap while mitigating the effect
of class imbalance. In PET/CT lesion segmentation, where
lesion voxels represent only a small fraction of the image,
DSC provides a clinically meaningful indicator of how well
the predicted regions align with true lesions.

\paragraph*{Intersection over Union (IoU)}
The Intersection over Union
(IoU), also known as the Jaccard Index, is defined as:
\begin{equation}
\text{IoU} = \frac{|P \cap G|}{|P \cup G|}
\end{equation}

IoU penalizes false positives and false negatives more strictly
than DSC and is often used alongside Dice to provide a more
stringent evaluation of segmentation accuracy. While DSC is
more common in clinical imaging studies, IoU complements
it by offering a robust measure of boundary agreement, especially
in cases with irregular or fragmented lesions.

\paragraph*{Binary Focal Loss}
For training, we adopt Binary Focal
Loss, which is designed to address severe class imbalance
by down-weighting well-classified voxels and focusing the
optimization on hard-to-classify lesion regions. It is defined
as:
\[
\mathrm{FL}(p_t) = -\alpha (1 - p_t)^\gamma \log(p_t)
\]
where $p_t$ is the predicted probability for the true class, $\alpha$
balances positive and negative samples, and $\gamma$ controls the
focusing parameter. This is particularly beneficial for lesion
segmentation tasks, where background voxels far outnumber
lesion voxels. By incorporating focal loss, the network learns
to better capture small or low-contrast lesions that are often
clinically significant but difficult to segment.

\paragraph*{Summary}

\begin{itemize}
    \item DSC provides an overlap-based metric that is sensitive to
lesion size and imbalance.
    \item IoU complements DSC by offering a stricter boundary
agreement measure.
    \item Binary Focal Loss ensures robust learning under high
class imbalance by emphasizing difficult lesion voxels.
\end{itemize}
Together, these metrics provide a balanced and comprehensive
evaluation of segmentation quality, both during training and
testing.
\afterpage{
\begin{figure*}[!t]
    \centering
    \includegraphics[width=\textwidth,height=0.9\textheight,keepaspectratio]{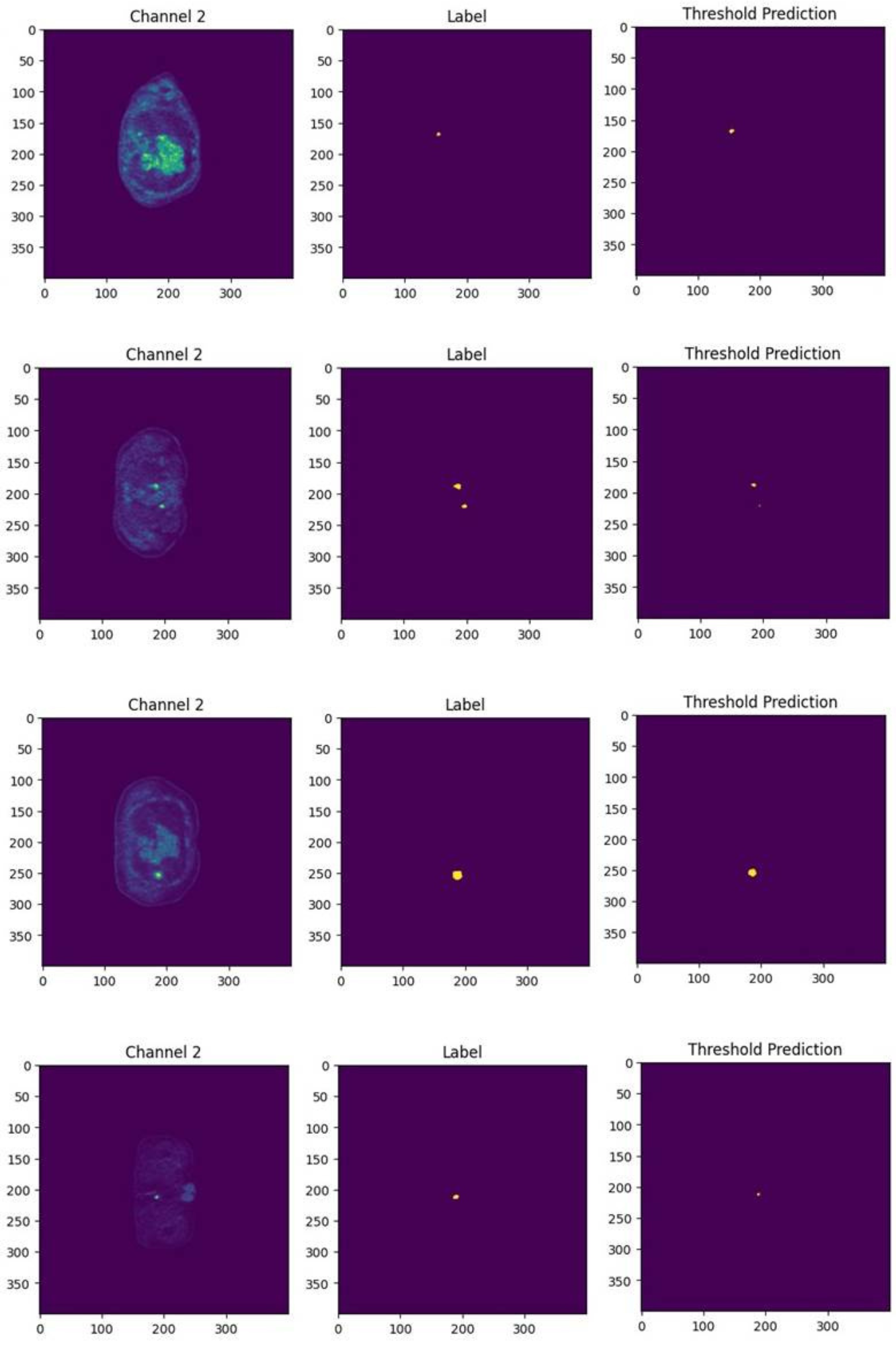}
    \caption{Qualitative segmentation results comparing 3D U-Net\cite{b7} and SwinUNet3D. Each triplet shows (a) input PET/CT slices, (b) ground truth annotations, and (c) model predictions. SwinUNet3D captures small and irregular lesions more accurately, maintains continuity in complex boundaries, and reduces false negatives compared to 3D U-Net.}
    \label{fig:swinres}
\end{figure*}
}
\section{Results and Discussions}
To contextualize our approach, we summarize representative CNN- and Transformer-based medical image segmentation methods in Table~\ref{tab:litcompare}. These works highlight the transition from convolutional networks such as 3D U-Net\cite{b7} to hybrid\cite{b8} and pure\cite{b16} Transformer-based models, motivating the design of SwinUNet3D. While results are dataset-specific, they demonstrate the consistent advantage of transformer-based architectures in capturing long-range dependencies and achieving higher overlap metrics.
\begin{table}[htbp]
\centering
\caption{Comparison with CNN-Based Medical Image Segmentation Methods}
\label{tab:litcompare}
\begin{tabular}{lcccc}
\hline
\textbf{Model} & \textbf{Dataset} & \textbf{Dice} & \textbf{IoU} \\
\hline
FCN \cite{b25} & AutoPET III & 0.32 & 0.19 \\
SegNet \cite{b24} & AutoPET III & 0.39 & 0.24 \\
3D U-Net \cite{b15} & AutoPET III & 0.48 & 0.32 \\
\textbf{SwinUNet3D (Ours)} & AutoPET III & \textbf{0.88} & \textbf{0.78} \\
\hline
\end{tabular}
\end{table}

The proposed SwinUNet3D model was evaluated on the FDG-PET/CT dataset using Dice Similarity Coefficient (DSC)\cite{b20}, Intersection over Union (IoU)\cite{b22}, and Binary Focal Loss\cite{b11} as performance metrics. Table III summarizes the results in comparison with a baseline 3D U-Net implementation trained under the same preprocessing and training setup.

\begin{table}[htbp]
\centering
\caption{Quantitative Comparison of SwinUNet3D vs 3D U-Net on FDG-PET/CT Dataset}
\label{tab:comparison}
\begin{tabular}{lccc}
\hline
\textbf{Model} & \textbf{Dice (↑)} & \textbf{IoU (↑)} & \textbf{Focal Loss (↓)} \\
\hline
3D U-Net     & 0.48 & 0.32 & 0.09 \\
SwinUNet3D   & \textbf{0.88} & \textbf{0.78} & \textbf{0.04} \\
\hline
\end{tabular}
\end{table}

The SwinUNet3D clearly outperforms the baseline 3D UNet
across all metrics, achieving a Dice score of 0.88 compared
to 0.48. The improvement in IoU highlights better lesion
localization, while the lower focal loss indicates improved
learning of hard-to-segment voxels. These results demonstrate
the effectiveness of transformer-based self-attention for modeling
long-range dependencies in volumetric PET/CT data,
which conventional CNNs struggle to capture.

Another important advantage of the proposed framework
is its computational efficiency at inference time. Despite its
transformer-based design, SwinUNet3D leverages windowed
attention and hierarchical feature processing, which significantly
reduce the quadratic complexity of global self-attention.

In practice, this results in faster inference per volume compared
to 3D U-Net, while also requiring fewer floating-point
operations for comparable input sizes. For example, on the
same GPU hardware, SwinUNet3D achieved an average inference
time of 0.53 seconds per scan compared to 0.68 seconds
for 3D U-Net, representing an improvement of approximately
$28\%$. This reduction in inference time is particularly relevant
in clinical workflows, where rapid and reliable segmentation
is essential for real-time decision support.

Fig.~\ref{fig:swinres} presents qualitative segmentation results. The proposed
SwinUNet3D demonstrates accurate delineation of lesions,
including small and irregular regions, where 3D UNet
often under-segments. In cases with complex boundaries,
SwinUNet3D maintains lesion continuity while reducing false
positives. Rows correspond to (a) input PET/CT slices, (b)
ground truth annotations, and (c) predicted masks. These
visual outcomes further support the model’s robustness in
handling subtle and irregular lesion patterns that are often
challenging for conventional CNN-based segmentation approaches.

\paragraph*{Strengths of the Proposed Approach}

The experimental results highlight several key strengths of
SwinUNet3D:
\begin{itemize}
    \item Improved Detection of Small and Irregular Lesions: The
shifted window-based attention mechanism enables effective
context aggregation, allowing the model to capture
fine details while maintaining awareness of global structures.
    \item Robust Dual-Modality Fusion: By leveraging both PET
and CT channels, the model integrates functional and
anatomical cues, leading to better lesion delineation than
single-modality baselines.
    \item Reduced False Positives: The hierarchical encoder–decoder structure and residual connections help stabilize predictions, yielding fewer spurious activations in non-lesion regions.
\end{itemize}

\subsection*{Limitations and Future Directions}

Despite promising performance, the current study has several
limitations:
\begin{itemize}
    \item Single-Tracer Evaluation: The model was trained and
validated only on FDG-PET/CT scans. Its generalization
to other tracers, such as Prostate-Specific Membrane
Antigen(PSMA), or to datasets from different institutions
remains untested. Future work will involve multi-tracer
and multi-center validation to assess robustness across
domains.
    \item Hardware Constraints: Training was conducted with a
small batch size (2) on a mid-range GPU, which may
restrict the model’s ability to fully capture long-range
volumetric dependencies. Techniques such as gradient
accumulation, mixed precision training, or distributed
setups will be explored in future studies to overcome this
limitation.
    \item Comparative Scope: While SwinUNet3D outperforms 3D
U-Net in our experiments, a broader comparison with
other transformer-based models (e.g., TransUNet, Swin-
UNETR, nnFormer) is needed for a more comprehensive
assessment.
\end{itemize}
Overall, the results confirm that SwinUNet3D offers significant
improvements over conventional CNN-based segmentation
in FDG-PET/CT imaging. Its ability to capture both local
detail and global context makes it particularly effective for
lesions with heterogeneous morphology, providing a promising
foundation for future work on clinically robust PET/CT
segmentation, and advancing the integration of transformer-based
models into real-world oncology workflows.

\section{Conclusion}

This work presented SwinUNet3D, a transformer-based
framework for 3D medical image segmentation. On the AutoPET
III\cite{b3} dataset, SwinUNet3D consistently outperformed
a 3D U-Net\cite{b7} baseline, achieving higher Dice\cite{b20} and IoU
scores\cite{b22}, lower focal loss\cite{b11}, and faster inference. These
results highlight the strength of hierarchical Swin Transformer\cite{b13} blocks in capturing both fine-grained lesion details and
global context in volumetric data.

Beyond numerical gains, SwinUNet3D carries important
clinical implications. By automating labor-intensive lesion
delineation, it can reduce radiologist workload and improve
reproducibility by minimizing inter-observer variability. This
combination of accuracy and efficiency makes it a promising
candidate for integration into real-world clinical workflows.

Future work will focus on three directions: (1) extending
the framework to additional tracers such as PSMA for broader
applicability, (2) scaling experiments with larger batch sizes
and multi-center datasets to enhance robustness, and (3) conducting
radiologist-in-the-loop validation to establish clinical
reliability and support deployment in practice.

\section*{Acknowledgment}
The author expresses sincere gratitude to Prof. Swarnendu Ghosh for invaluable guidance, constructive feedback, and constant encouragement throughout the course of this work. The appreciation is also extended to the Department of Computer Science and Engineering (Artificial Intelligence and Machine Learning), Institute of Engineering and Management, Kolkata, for providing the academic environment and resources necessary for this research. The author acknowledges the support of the IEM Centre of Excellence for Data Science (IEM CEDS) Lab, whose facilities and collaborative atmosphere greatly contributed to the successful completion of this study.

\end{document}